\documentstyle[epsf,psfig,graphicx]{mn}

\def\x2{$\chi^{2}$}

\def\lunits{$\rm erg~s^{-1}$}
\def\funits{$\rm erg~cm^{-2}~s^{-1}$}
\def\cunits{$\rm cm^{-2}$}

\newbox\grsign \setbox\grsign=\hbox{$>$} \newdimen\grdimen \grdimen=\ht\grsign
\newbox\simlessbox \newbox\simgreatbox \newbox\simpropbox
\setbox\simgreatbox=\hbox{\raise.5ex\hbox{$>$}\llap
     {\lower.5ex\hbox{$\sim$}}}\ht1=\grdimen\dp1=0pt
\setbox\simlessbox=\hbox{\raise.5ex\hbox{$<$}\llap
     {\lower.5ex\hbox{$\sim$}}}\ht2=\grdimen\dp2=0pt
\setbox\simpropbox=\hbox{\raise.5ex\hbox{$\propto$}\llap
     {\lower.5ex\hbox{$\sim$}}}\ht2=\grdimen\dp2=0pt

\def\asca{{\it ASCA~}}
\def\sax{{\it BeppoSAX}}
\def\rosat{{\it ROSAT~}}

\def\xmm{{\it XMM-Newton~}}
\def\chandra{{\it Chandra~}}

\begin{document}

\title[X-ray spectral properties of the hard X-ray sources]
{The XMM-Newton/2dF survey IV. The X-ray spectral properties 
of the hard sources}

\author[Georgantopoulos et al.]
{ I. Georgantopoulos$^1$, A. Georgakakis$^1$, 
  A. Akylas $^{1,2}$, G.C. Stewart$^3$,  O. Giannakis$^{1,2}$, \\ \\
  {\LARGE  T. Shanks$^4$, S. Kitsionas$^1$} \\
$^1$ Institute of Astronomy \& Astrophysics, National Observatory of Athens, 
 I. Metaxa \& B. Pavlou, Penteli, 15236, Athens, Greece \\
$^2$ Physics Department University of Athens, Panepistimiopolis, 
 Zografos, 15783, Athens, Greece \\
$^3$ Department of Physics and Astronomy, University of Leicester, 
 Leicester LE1 7RH \\  
$^4$ Physics Department, South Road, Durham, DH1 3LE  \\ 
}

\maketitle
\begin{abstract}
 We present an analysis of the X-ray spectral properties of 61 hard
 X-ray selected (2-8\,keV) sources from the bright 
 ($f_{2-8 keV}>10^{-14}$ \funits)  {\it XMM-Newton}/2dF
 survey. This comprises of 9 {\it XMM-Newton} pointings in the North Galactic
 Pole region ($\sim 1.6 \rm deg^2$) and overlaps with the SDSS, 2QZ and
 2dFGRS surveys. Our sources contribute about 50 per cent of the  
 2-10\,keV X-ray background down to the flux limit of $10^{-14}$
 \funits. The hardness ratio distribution of the sample suggests a
 deficit of heavily absorbed sources.  
 A spectral fit to the co-added total source spectrum yields  
 a steep photon index, $\Gamma=1.83^{+0.04}_{-0.05}$.
 All but 8 sources have optical counterparts down to
 the  SDSS photometric limit of $r\approx22.5$. Spectroscopic
 identifications exist for 34 sources. The vast majority are
 associated with Broad-Line (BL) AGN (24 sources) while only 
 7 present narrow or no emission lines. Five sources are 
 probably associated with Galactic stars.  
 Finally, for another 17 probable   AGN   
 we present photometric redshifts. 
 The combined spectrum of the 24 spectroscopically identified 
 BL AGN is steep ($\Gamma=2.02^{+0.04}_{-0.05}$), while that  
 of the 7 AGN, which do not present 
 broad lines is flatter with $\Gamma=1.64^{+0.11}_{-0.11}$. 
 The spectrum of the 8 optically unidentified sources is flat with  
 $\Gamma\approx 1.1$.  Spectral fits to the individual
 BL AGN reveal large absorption (rest-frame column density
 $\rm>10^{22}\,cm^{-2}$) in only two cases.
 The individual spectra of the NL AGN present significant 
 evidence for even a moderate 
 absorption ($\rm 3\times 10^{21}\,cm^{-2}$) in only one case. 
      
\end{abstract}


\section{Introduction}
 Deep \chandra surveys have resolved the bulk of the X-ray background
 in both the soft and the hard energies  (Mushotzky et al. 2000;
 Brandt et al. 2001; Giaconni et al. 2002, Alexander et al. 2003). 
 These surveys detect a heterogeneous population of sources  
 consisting of a mixture of
 (i) BL AGN  (QSOs and Seyfert-1 galaxies), (ii) narrow
 emission line  AGN, (iii) 'passive' galaxies with absorption line
 optical spectra and (iv) optically faint sources
 ($I>24$\,mag). It is likely, that both the 'passive' galaxies and the 
 optically faint sources may be associated 
 with obscured AGN (see e.g. Moran et al. 2002,
 Fiore et al. 2003). 
 Nevertheless, deep \chandra observations primarily probe the faint
 ($<10^{-16}$ \funits) X-ray population. At brighter fluxes
 ($>10^{-15}$ \funits) \chandra is limited by its small field of view
 ($0.07\, \rm deg^2$) and cannot provide large 
 samples. However, it is these bright sources that dominate the X-ray 
 Background (XRB)
 ($>50$ per cent , e.g. Boyle et al. 1998). Therefore, to
 improve our understanding of the sources that make up the XRB, wide
 area relatively shallow surveys are essential.  

 \xmm with its large field-of-view, high effective area 
 and good positional accuracy provides a step forward in
 the study of these sources.  In this paper, we present the X-ray
 spectral properties of sources detected in a shallow \xmm 
 survey covering  a $\sim 1.6 \rm deg^2$ area near the North Galactic Pole
 region. In particular, we concentrate on the hard X-ray selected sample 
 (2-8\,keV) as these are more typical  
 of the sources contributing to the XRB with its energy density
 peaking at 30-40\,keV.  Although extensive surveys have been
 conducted at these energies  by both the \asca and \sax \ missions
 (e.g. Georgantopoulos et al.  1997, Ueda et al. 1998, Giommi et
 al. 2000), the poor spatial resolution  hampered the identification
 of the optical counterparts.  The present data probe a factor
 of $\approx4$ fainter fluxes than the deepest \asca survey
 (Georgantopoulos et al. 1997) with an order of magnitude better
 positional accuracy.   
   
 Throughout this paper we adopt $\rm H_o=65 km~s^{-1}~Mpc^{-1}$ and
 $q_o=0.3, \Lambda=0.7$.   
    
\section{XMM-Newton/2dF survey overview}
The X-ray sample used in the present study is compiled from the
{\it XMM-Newton}/2dF survey. This is a wide area ($\rm \sim 4\,deg^2$) shallow
(2-10\,ks per pointing) survey carried out by the \xmm near the
North and the South Galactic Pole regions. The data reduction, source
extraction, flux estimation  and catalogue generation are described in
detail by Georgakakis et al. (2003, 2004). In the present study we
concentrate on the North Galactic Pole F864  region. This is because
of the  wealth of follow-up observations (optical photometry and
spectroscopy)  available for these fields. 

The {\it XMM-Newton}/2dF survey F864 region overlaps with
the Sloan Digital Sky Survey, SDSS, (York et al. 2000). The  SDSS is an
on-going imaging and spectroscopic survey that aims to cover about $\rm
10\,000\,deg^2$ of the sky. Photometry is performed in 5 bands
($ugriz$;  Fukugita et al. 1996; Stoughton et al. 2002) to the
limiting magnitude $g \approx 23$\,mag, providing a uniform and
homogeneous multi-color photometric catalogue. The SDSS spectroscopic
observations will obtain spectra for over 1 million objects, including
galaxies brighter than $r=17.7$\,mag, luminous red galaxies to
$z\approx0.45$ and colour selected QSOs (York et al. 2000; Stoughton
et al. 2002). In the present study we use data from the Early Data
Release (EDR; Stoughton et al. 2002).

In addition to the SDSS the F864 region overlaps with the
recently completed 2dF Galaxy Redshift Survey
(2dFGRS\footnote{http://msowww.anu.edu.au/2dFGRS/}; Colless et
al. 2001; Colless et al. 2003) and the 2dF QSO Redshift Survey
(2QZ\footnote{http://www.2dfquasar.org}; Croom et al. 2001). Both the
2dFGRS and 2QZ are large-scale spectroscopic campaigns that fully
exploit the capabilities of the 2dF multi-fibre spectrograph on the
4\,m Anglo-Australian Telescope (AAT). These
projects provide high quality spectra, redshifts and spectral
classifications for 220\,000 $bj<19.4$\,mag galaxies and 23\,000
optically selected $bj<20.85$\,mag QSOs. Moreover, the central region
of the F864 survey have been observed by
the {\it ROSAT} satellite (Shanks et al. 1991, Georgantopoulos et
al. 1996). 

\begin{table*} 
\footnotesize 
\begin{center} 
\begin{tabular}{ccccc} 
\hline 
Field Name & RA          & Dec       & PN exp. time & MOS exp. time \\  
           & (J2000)   & (J2000) &     (sec) &  (sec) \\
\hline 
F864-1
& $13\mathrm{^h} 41\mathrm{^m} 24.0\mathrm{^s}$ 
& $+00\mathrm{^\circ} 24\mathrm{^\prime} 00\mathrm{^{\prime\prime}}$ 
& 5779  & 9974 \\
F864-2
& $13\mathrm{^h} 43\mathrm{^m} 00.0\mathrm{^s}$ 
& $+00\mathrm{^\circ} 24\mathrm{^\prime} 00\mathrm{^{\prime\prime}}$ 
& 2958  & 6586 \\
F864-3
& $13\mathrm{^h} 44\mathrm{^m} 36.0\mathrm{^s}$ 
& $+00\mathrm{^\circ} 24\mathrm{^\prime} 00\mathrm{^{\prime\prime}}$ 
& 2187  & 7727 \\
F864-4  
& $13\mathrm{^h} 41\mathrm{^m} 24.0\mathrm{^s}$ 
& $+00\mathrm{^\circ} 00\mathrm{^\prime} 00\mathrm{^{\prime\prime}}$ 
& -- & -- \\
F864-5
& $13\mathrm{^h} 43\mathrm{^m} 00.0\mathrm{^s}$ 
& $+00\mathrm{^\circ} 00\mathrm{^\prime} 00\mathrm{^{\prime\prime}}$ 
& 1693 & 4447 \\
F864-6
& $13\mathrm{^h} 44\mathrm{^m} 36.0\mathrm{^s}$ 
& $+00\mathrm{^\circ} 00\mathrm{^\prime} 00\mathrm{^{\prime\prime}}$ 
& 2766 & 6493 \\
F864-7
& $13\mathrm{^h}$ $41\mathrm{^m} 24.0\mathrm{^s}$ 
& $-00\mathrm{^\circ} 24\mathrm{^\prime} 00\mathrm{^{\prime\prime}}$ 
& 3459 & 7139 \\
F864-8
& $13\mathrm{^h} 43\mathrm{^m} 24.0\mathrm{^s}$ 
& $-00\mathrm{^\circ} 24\mathrm{^\prime} 00\mathrm{^{\prime\prime}}$ 
& 2109 & 7276 \\
F864-9
& $13\mathrm{^h} 44\mathrm{^m} 36.0\mathrm{^s}$ 
& $-00\mathrm{^\circ} 24\mathrm{^\prime} 00\mathrm{^{\prime\prime}}$ 
& 4545 & 8330 \\	

\hline 

\end{tabular} 
\end{center} 
\caption{Observing log  of the {\it XMM-Newton}/2dF survey F864
region}
\normalsize  
\label{log}
\end{table*} 

\section{The X-ray data}
The EPIC (European Photon Imaging Camera; Str\"uder et al. 2001 and
Turner et al. 2001) cameras were operated in full frame mode with the
thin filter applied. The {\it XMM-Newton} data have been analysed
using the Science Analysis Software (SAS 5.3). Event files for the PN
and the two MOS detectors have been produced using  the {\sc epchain}
and {\sc emchain} tasks of SAS respectively. The event files were
screened for high particle  background periods by rejecting periods with
0.5-10\,keV count rates higher than 25 and 15\,cts/s for the PN and
the two MOS cameras respectively. These criteria exclude one 
field (F864--4) from the analysis that suffered by significantly
elevated and flaring particle background. The resulting 
 area covered by the survey is $1.6\rm deg^2$. 
The PN and MOS good time
intervals for the remaining pointings are shown in Table 
\ref{log}. The differences between the PN and MOS exposure times are
due to varying start and end times of individual observations. Only
events  corresponding to patterns  0--4 for the PN and 0--12 for 
MOS have been kept. 

In order to increase the signal--to--noise ratio and to reach fainter fluxes
the PN and the MOS event files have been combined into a single event
list using the {\sc merge} task of SAS. Images have been extracted in the
spectral bands 0.5-8 (total), 0.5-2 (soft) and 2-8\,keV (hard) for
both the merged and the individual PN and MOS event files. We use the 
more sensitive (higher S/N ratio) merged image for source  extraction
and flux estimation, while the individual PN and MOS images are used
to calculate hardness ratios. This is because the interpretation of
hardness ratios is simplified if the extracted count rates  are from
one detector only.  Exposure maps accounting for vignetting, CCD gaps
and bad pixels  have been constructed for each spectral band. In the
present study the source detection is performed on the 2-8\,keV image
using the {\sc ewavelet} task of SAS with a detection threshold of 
$6\sigma$. A total of 61 X-ray sources have been detected to the limit
$f_X(\rm 2-8\,keV)\approx10^{-14}\,erg\,s^{-1}\,cm^{-2}$. 
Integration of the logN-logS derived by Baldi et al. (2002) to 
 the above flux limit gives, shows that about 50 per cent of the 
 XRB is resolved at these energies.  

Count rates in the merged (PN+MOS) images as well as the
individual PN and MOS images are estimated within an 18\,arcsec
aperture. For the  background estimation we use the background maps
generated as a by-product of the {\sc ewavelet} task of  SAS. A small
fraction of sources lie close to masked regions (CCD gaps or hot
pixels) on either the MOS or the PN detectors. This may introduce
errors in the estimated source  counts. To avoid this bias, the source
count rates (and hence the hardness ratios and the flux) are estimated
using the detector (MOS or PN) with no masked pixels in the vicinity
of the source. 

We convert counts to flux assuming a power-law spectrum with
$\Gamma=1.7$ and Galactic absorption $N_H=2\times 10^{20} \rm
{cm^{-2}}$ appropriate for  the F864 fields (Dickey \& Lockman
1990). Note that a $\Gamma=1.7$ photon index 
 provides a reasonable representation of the typical source spectrum
 being intermediate between the steep BL AGN and the unidentified flatter 
 sources (see section 5.3).  
 The mean ECF for  the mosaic of all three detectors is
estimated by weighting the ECFs of individual detectors by the
respective exposure time.  For the encircled energy correction, 
accounting for the energy fraction outside the aperture within which
source counts are accumulated, we adopt the calibration 
 given by the {\it XMM-Newton} Calibration Documentation
\footnote{http://xmm.vilspa.esa.es/external/xmm\_sw\_cal/calib \\
/documentation.shtml\#XRT}.
 These studies use both PN and MOS
observations of point sources to formulate the \xmm PSF  
for different energies and off-axis angles. In particular, a King
profile is fit to the data with parameters that are a function
of both energy and off-axis angle.

\section{The Optical Counterparts} 

The SDSS optical photometric catalogue is used to optically identify
the hard selected sample by estimating the probability, $P$, a given
candidate is chance coincidence (Downes et al. 1986). 
Of the 61 hard X-ray sources we propose 49
candidate optical identifications with  $P<0.015$;
 8 sources have no optical counterpart.
 4 sources have a probability higher than 0.015 of being 
 spurious and thus they are less secure identifications.  
 The probability depends on both the separation of the 
 optical counterpart from the X-ray centroid and  
 the surface density of the optical sources   
 at the given magnitude (see Georgakakis et al. 2004 for details). 
 To estimate the fraction of spurious identifications 
 in our sample we randomise the X-ray
 source positions and re-estimate the number of optical
 counterparts. This is repeated 100 times. We find a spurious
 identification  rate of $\approx6$ per cent or about 4 spurious 
 optical counterparts out of the 61 sources. 

 The catalogue of the X-ray sources is given in Table \ref{master}. 
 Optical spectroscopic information are available for 34 out of the 61 
 sources.  These contribute 
 about 30 per cent of the XRB in the 2-8 keV energy band. 
  A total of 21 sources  have redshifts and optical 
 classifications from our own
 spectroscopic campaign using the 2dF (Georgantopoulos et al. 2004
 in preparation). The remaining sources have spectroscopic data 
 from the 2QZ, 2dFGRS and the SDSS surveys. 
 The spectroscopically identified sample comprises: 
 (i) 24 sources with broad optical emission lines (BL AGN). 
 (ii) five sources with  
 narrow lines (NL) (iii) two  with absorption lines (AL) 
 As all the NL/AL objects have luminosities
  in excess of $10^{42}$ \lunits, these are most probably AGN.
 (iv) three stars; note that closer inspection of the 
 (low signal-to-noise)  spectrum of source \#2 
 which is classified 
 as a star by the 2QZ survey, leaves open the possibility 
 that this is associated with a galaxy at a redshift of z=0.243.    
  We also note that two additional sources 
 for which no spectroscopic information is available (\#1 and \#26)
 are candidate stars on the basis of their low $f_x/f_o$ ratio
 and optical colours (for the latter see e.g. 
 Hatziminaoglou, Mathez \& Pello 2000).  
 In addition, to the spectroscopic
 refshifts, we have estimated photometric redshifts, using the
 multiwaveband coverage of the SDSS, for 6 X-ray sources 
 which are most probably QSOs as they have 
 point-like optical profile and high X-ray to optical flux ratio. 
 The photometric redshift technique is fully described in 
 Kitsionas et al. (in preparation) and Hatziminaoglou et al. (2000).
 Note that point-like sources with a determined photometric redshift of 
 $z<0.4$, are likely to present large uncertainties in their
 redshift (Kitsionas et al. in preparation).  
 For 11 sources (\#4, 6, 13, 14, 15, 18, 22, 24, 46, 47, 60) 
 which are optically extended (most have also red colours), 
 we have used the photometric redshifts provided by SDSS which are 
 based on galaxy templates (Csabai et al. 2003).  

\begin{figure}
\rotatebox{270}{\epsfxsize=6.5cm \epsffile{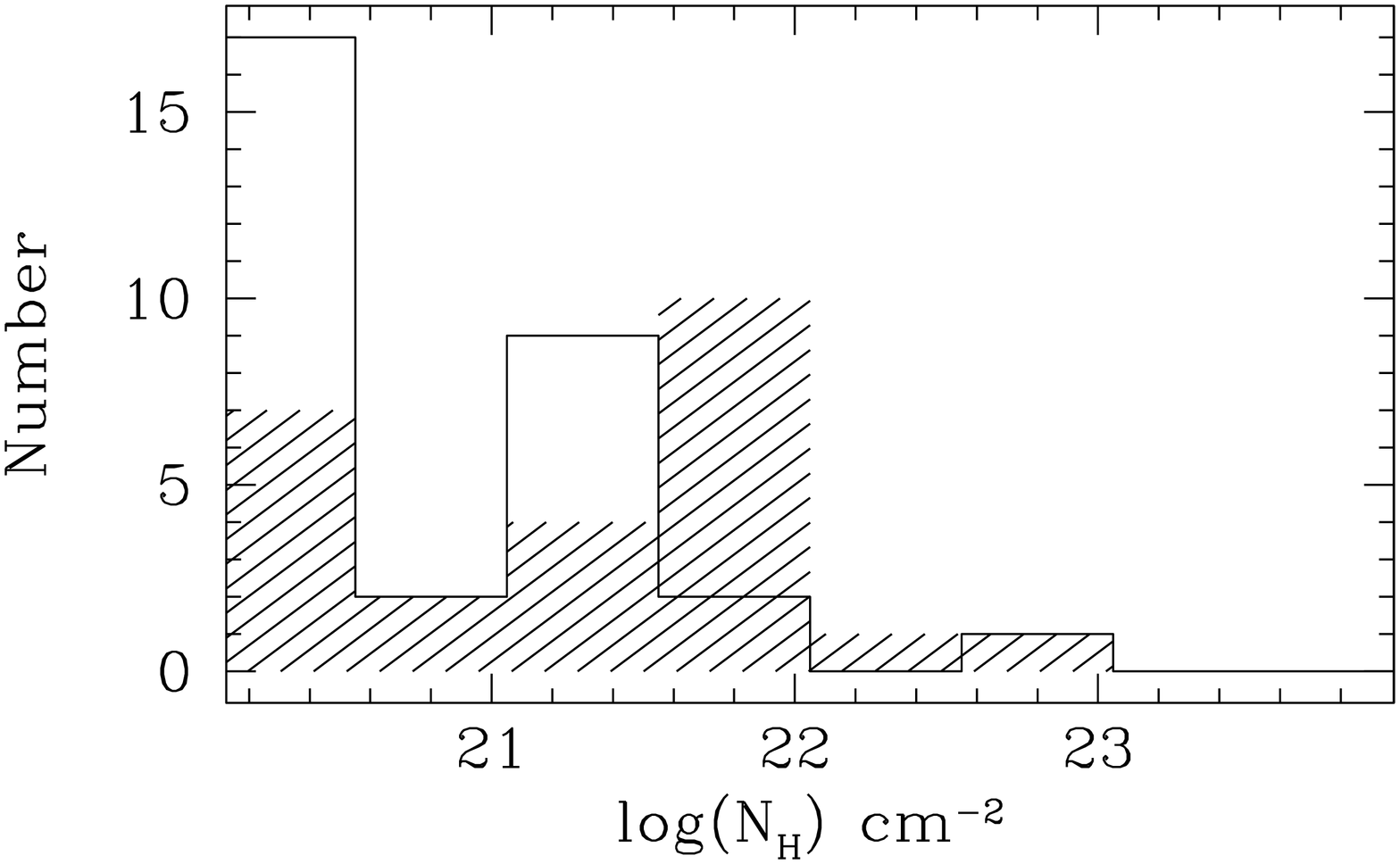}}   
\caption{The {\it observer's frame} column density distribution
 for the spectroscopically unidentified (open) and identified 
 (hatched) samples. 
 }
 \label{nh}
\end{figure}

\begin{figure}
\rotatebox{270}{\epsfxsize=6.0cm \epsffile{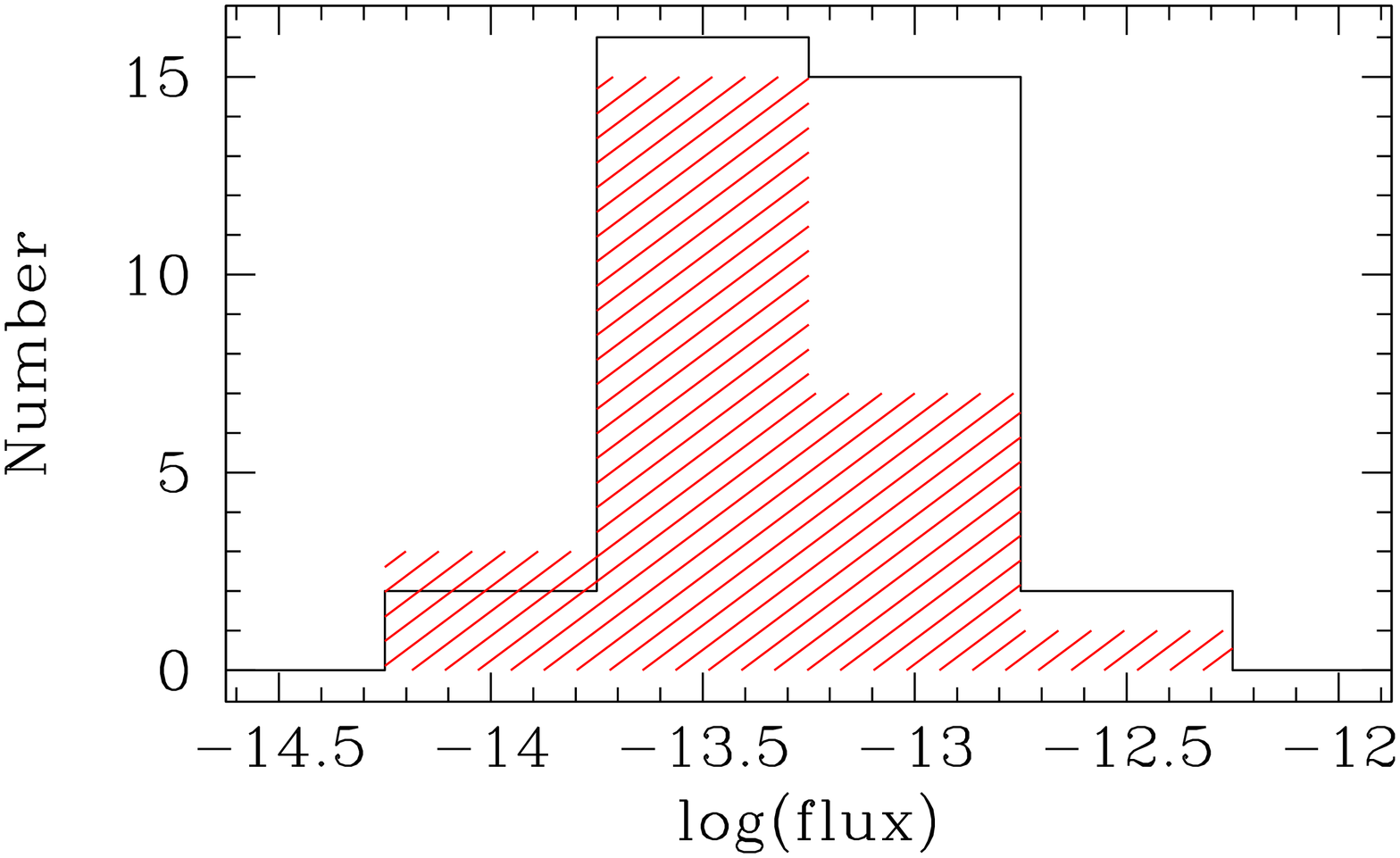}}   
\caption{Flux (2-8 keV) distribution of the spectroscopically identified 
 (hatched) and unidentified (open) sources.  
 }
 \label{fx}
\end{figure}

 As a fraction of the spectroscopic identifications come 
 from the 2QZ survey, there may be a bias towards the 
 detection of a high number of QSOs among our identified sources.
 We plot the column density and flux distributions, in order to 
 further understand any possible biases introduced.
 The {\it observer's frame} column density is derived from the
 hardness ratios as described in section 5.1.      
 Indeed, from Fig. \ref{nh}, we see that the unidentified sources are 
 in general harder. This is
 not surprising as many of our optical identification are  based on the
 SDSS and 2QZ surveys which are designed to detect  UV excess selected QSOs
 most of which do not show strong  absorption in X-rays. 
 In Fig. \ref{fx}, we present the flux distribution,
 again separating the spectroscopically identified 
 and the unidentified sources. 
 Both the spectroscopically identified and unidentified sources 
 appear to sample the whole flux range. 
 However, at the faintest flux bin, there are more unidentified sources. 

\begin{figure*}
\rotatebox{270}{\epsfxsize=9.0cm \epsffile{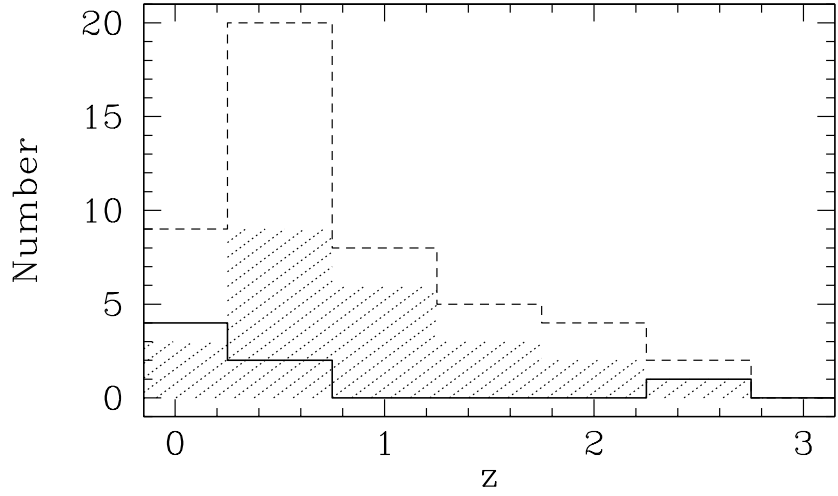}}   
\caption{The redshift distributions of the spectroscopically 
 identified BL AGNs (hatched histogram),  NL/AL AGN 
 (solid line) and the full sample i.e. including  
 photometric redshifts (dashed line).
 }
 \label{nz}
\end{figure*}

\begin{figure*}
\rotatebox{270}{\epsfxsize=10.0cm \epsffile{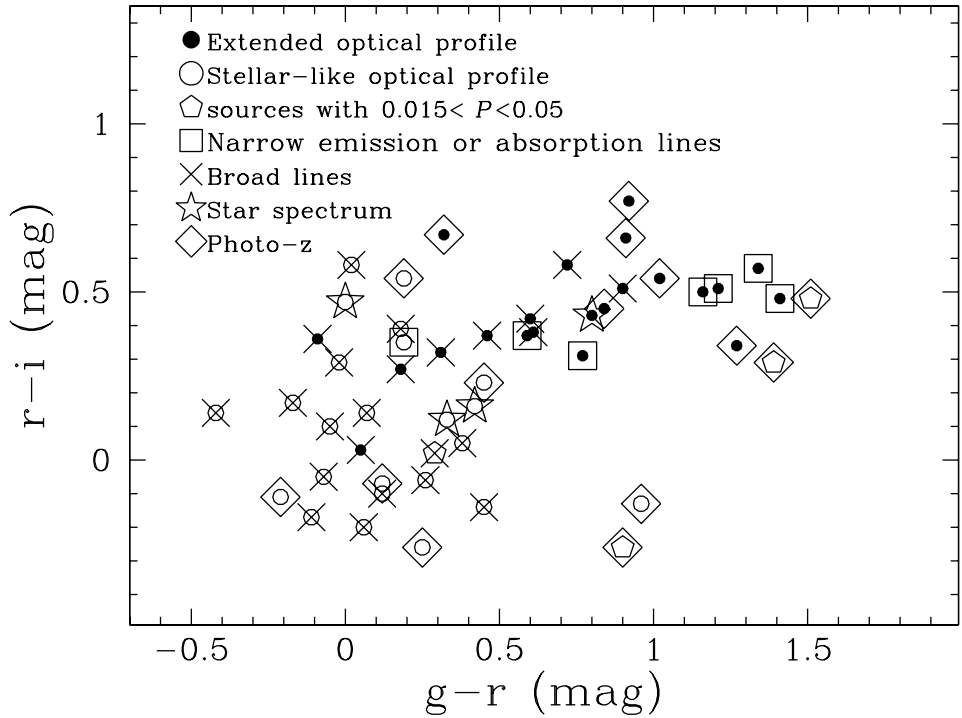}}   
\caption{
 $g-r$ against $r-i$ colour-colour diagram for the hard X-ray 
selected sample.  
 Filled and open circles on top of a symbol correspond to the sources
 with stellar-like and extended optical morphology according to the
 SDSS. The pentagon is for sources with higher probability of being
 spurious alignments ($0.015<P<0.05$).  Open squares are for NL/AL, 
 crosses are BL AGNs and
 stars correspond to spectroscopically confirmed stars. The diamonds 
 correspond to sources with photometric redshift estimates. 
 }\label{colour}
\end{figure*}

In Fig. \ref{nz} we plot the redshift distribution of the 
 BL (24) and AL/NL (7) AGN. The NL AGN are preferably detected  
 at low redshift in agreement with the findings of other 
 hard X-ray surveys (e.g. Brusa et al. 2003).  
Fig. \ref{colour} shows the colour-colour plot for the optically
identified hard X-ray sources. A striking result from this Figure is
the significant number of red ($g-r>0.5$) sources with extended
optical profiles suggesting relatively low-$z$ galaxies. Fig.
\ref{colour} suggests that these sources are a non-negligible mode of
the hard X-ray selected population to the limit of our survey.

\begin{table*}
\caption{Optical and X-ray properties of the sample}
\label{master}
\begin{tabular}{ccc ccc ccc ccc}
\hline
\# & $\alpha_X$ & $\delta_X$ & count rate$^a$ & flux$^b$ & exp. time$^c$ & $P$ & $\delta r$ & $g$ & optical & $z$ & class \\
  &(J2000) & (J2000) &        &       &   &  ($\%$) & (arcsec) & (mag) & class &  & \\
(1) &  (2) & (3) & (4) & (5) & (6) & (7) & (8) & (9) & (10) & (11) & (12) \\    
\hline
1$^{}$ & 13 45 15.0  & --00 00 48  &  $8\pm3.9$ & $9.78\pm2.04$ & 6871 & $<0.01$ & 0.3 & 15.44 & point-like & 0.000 &STAR$^{f}$\\
2$^{}$ & 13 45 10.4  & +00 18 51  &  $6.8\pm4$ & $5.14\pm1.43$ & 9717 & 0.19 & 1.6 & 19.17 & extended & 0.000 &STAR?$^g$\\
3$^{}$ & 13 45 09.8  & +00 20 52  &  $11.1\pm4.49$ & $15.7\pm2.32$ & 8137 & 0.82 & 2.4 & 19.54 & extended & 0.370 &NL\\
4$^{}$ & 13 45 08.5  & +00 22 27  &  $1\pm2.6$ & $2.52\pm1.06$ & 11523 & 1.19 & 1.9 & 21.05 & extended & 0.415 &photo-z$^{e}$\\
5$^{}$ & 13 45 08.0  & --00 05 29  &  $12.7\pm4.6$ & $10.3\pm2.05$ & 7183 & 0.26 & 2.2 & 19.77 & extended & 0.730 &BL\\
6$^{}$ & 13 45 07.4  & +00 04 09  &  $6.7\pm3.4$ & $5.66\pm1.56$ & 7724 & 0.67 & 2.6 & 19.52 & extended & 0.553 &photo-z$^{e}$\\
7$^{}$ & 13 45 01.7  & --00 24 02  &  $5.5\pm1.9$ & $3.77\pm0.83$ & 16074 & 0.01 & 0.8 & 17.38 & point-like & 0.000 &STAR\\
8$^{d}$ & 13 44 59.5  & --00 16 01  &  $14.7\pm1.5$ & $27.5\pm2.89$ & 9824 & 0.05 & 1.8 & 17.52 & point-like & 0.245 &BL\\
9$^{}$ & 13 44 58.5  & +00 16 23  &  $5.1\pm3.6$ & $3.13\pm1.21$ & 10167 & 0.26 & 2.9 & 18.02 & extended & 0.145 &NL\\
10$^{}$ & 13 44 58.0  & --00 36 00  &  $12.5\pm3.8$ & $8.69\pm1.64$ & 8110 & 1.36 & 4.4 & 19.38 & extended & 0.465 &BL\\
11$^{d}$ & 13 44 52.9  & +00 05 21  &  $20.2\pm1.9$ & $37.8\pm3.54$ & 9789 & 0.01 & 0.7 & 16.32 & extended & 0.087 &BL\\
12$^{}$ & 13 44 52.3  & --00 36 54  &  $4.2\pm2.7$ & $2.78\pm1.8$ & 5895 & 0.77 & 3.3 & 20.3 & point-like & 0.810 &photo-z\\
13$^{}$ & 13 44 51.6  & --00 23 01  &  $4.49\pm1.6$ & $2.5\pm0.65$ & 18932 & 0.30 & 1.4 & 19.64 & extended & 0.261 &photo-z$^{e}$\\
14$^{}$ & 13 44 47.0  & --00 30 09  &  $3\pm1.6$ & $2.45\pm0.74$ & 15454 & 0.67 & 1.1 & 20.18 & extended & 0.529 &photo-z$^{e}$\\
15$^{}$ & 13 44 44.1  & --00 19 30  &  $2\pm1.3$ & $1.67\pm0.62$ & 17078 & 4.87 & 2.6 & 22.28 & extended & 0.779 &photo-z$^e$\\
16$^{d}$ & 13 44 38.9  & --00 11 01  &  $5.7\pm1.19$ & $10.6\pm2.32$ & 6173 & -- & - & -- & -- & -- &--\\
17$^{}$ & 13 44 36.4  & +00 33 24  &  $11.1\pm4.49$ & $4.71\pm1.45$ & 9018 & 1.38 & 3.7 & 20.16 & point-like & 1.430 &BL\\
18$^{}$ & 13 44 33.5  & --00 24 55  &  $4.1\pm1.4$ & $3.35\pm0.69$ & 20492 & 0.32 & 0.8 & 22.91 & extended & 0.511 &photo-z$^{e}$\\
19$^{}$ & 13 44 25.3  & --00 18 27  &  $5.2\pm1.8$ & $3.7\pm0.85$ & 14566 & 0.34 & 1.5 & 21.71 & point-like & 1.970 &BL\\
20$^{d}$ & 13 44 24.7  & --00 13 08  &  $2.8\pm0.9$ & $5.26\pm1.65$ & 7915 & 2.19 & 4.7 & 20.41 & point-like & 1.110 &BL\\
21$^{d}$ & 13 44 22.1  & --00 34 20  &  $2.9\pm0.9$ & $5.36\pm1.61$ & 8539 & 0.10 & 1.3 & 18.2 & extended & 0.217 &NL\\
22$^{}$ & 13 44 20.2  & +00 04 17  &  $5.4\pm2.6$ & $4.76\pm1.09$ & 12779 & 2.45 & 3.3  & 19.98 & extended & 0.302 & photo-z$^e$ \\
23$^{}$ & 13 44 20.1  & --00 31 11  &  $5\pm2.1$ & $4.56\pm1.08$ & 11903 & 0.02 & 0.5 & 20.33 & extended & 0.680 &BL\\
24$^{d}$ & 13 44 14.2  & +00 16 41  &  $29.8\pm2.39$ & $55.8\pm4.42$ & 8546 & 0.06 & 1.1 & 19.34 & extended & 0.706 &photo-z$^{e}$\\
25$^{}$ & 13 44 04.7  & --00 09 23  &  $7.8\pm4.4$ & $6.96\pm1.83$ & 6549 & -- & --  & -- & --  & -- &--\\
26$^{}$ & 13 43 52.5  & --00 04 33  &  $14.4\pm5.2$ & $8.95\pm1.91$ & 7327 & 0.97 & 8.2 & 16.99 & point-like & -0.000 &STAR$^{f}$\\
27$^{d}$ & 13 43 51.1  & +00 04 38  &  $3.3\pm1$ & $6.25\pm1.84$ & 6311 & 0.16 & 3.4 & 16.75 & extended & 0.074 &BL\\
28$^{d}$ & 13 43 47.5  & +00 20 21  &  $6.6\pm1.3$ & $12.4\pm2.44$ & 5940 & 0.37 & 2.6 & 18.14 & extended & 0.240 &AL\\
29$^{d}$ & 13 43 31.6  & +00 24 49  &  $2.39\pm0.7$ & $4.56\pm1.31$ & 9978 & 0.82 & 3.4 & 20.27 & point-like & 1.300 &photo-z\\
30$^{d}$ & 13 43 29.2  & +00 01 35  &  $3.7\pm1.5$ & $6.94\pm2.85$ & 5307 & 0.75 & 2.2 & 21.04 & point-like & 2.347 &NL\\
31$^{}$ & 13 43 23.8  & +00 12 21  &  $8.4\pm4$ & $9.13\pm1.93$ & 6390 & 0.26 & 3.3 & 18.31 & point-like & 0.874 &BL\\
32$^{}$ & 13 43 07.9  & +00 27 19  &  $2.2\pm1.6$ & $2.21\pm0.73$ & 13999 & 0.42 & 2.4 & 20.87 & point-like & 2.310 &BL\\
33$^{}$ & 13 43 01.5  & +00 26 34  &  $5.6\pm2$ & $4.94\pm0.95$ & 14735 & 0.75 & 2.2 & 20.42 & point-like & 1.500 &photo-z\\
34$^{d}$ & 13 42 56.6  & +00 00 57  &  $8.1\pm1.3$ & $15.2\pm2.49$ & 10078 & 0.06 & 1.3 & 18.77 & point-like & 0.804 &BL\\
35$^{}$ & 13 42 55.5  & +00 06 36  &  $11.2\pm4.4$ & $7.87\pm2.08$ & 7602 & 0.13 & 1.5 & 19.33 & extended & 0.436 &BL\\
36$^{d}$ & 13 42 46.3  & --00 35 44  &  $6\pm1.7$ & $11.3\pm3.19$ & 6054 & 0.01 & 0.5 & 18.14 & point-like & 0.787 &BL\\
37$^{}$ & 13 42 42.8  & +00 32 23  &  $7.2\pm3.2$ & $4.74\pm2.11$ & 7858 & -- & - & -- & -- & -- &--\\
\hline
\multicolumn{12}{l}{$^a$ 2-8\,keV count rate in $\rm cts\ s^{-1}$}\\
\multicolumn{12}{l}{$^b$ 2-8\,keV flux in units of $\rm 10^{-14}\,erg\,s^{-1}\,cm^{-2}$}\\
\multicolumn{12}{l}{$^c$ total exposure time in the 2-8\,keV  band from the merged PN+MOS image in seconds}\\
\multicolumn{12}{l}{$^d$ count rate and flux is from MOS}\\
\multicolumn{12}{l}{$^e$ photometric redshifts are from the SDSS catalogue}\\
\multicolumn{12}{l}{$^f$ X-ray/optical properties indicate Galactic star but spectroscopy is not available}\\
\multicolumn{12}{l}{$^g$ The identification from 2QZ gives a star. However, the spectrum 
 may be suggesting a galaxy at z=0.243}\\
\hline
\end{tabular}
\end{table*}

\begin{table*}
\contcaption{}
\begin{tabular}{ccc ccc ccc ccc}
\hline
\# & $\alpha_X$ & $\delta_X$ & count rate$^a$ & flux$^b$ & exp. time$^c$ & $P$ & $\delta r$ & $g$ & optical & $z$ & class \\
  &(J2000) & (J2000) &        &       &   &  ($\%$) & (arcsec) & (mag) & class &  & \\
(1) &  (2) & (3) & (4) & (5) & (6) & (7) & (8) & (9) & (10) & (11) & (12) \\   
\hline
38$^{}$ & 13 42 34.5  & +00 32 23  &  $2.9\pm2.7$ & $4.02\pm1.43$ & 6883 & 1.32 & 2.4 & 20.98 & point-like & 1.950 &photo-z\\
39$^{d}$ & 13 42 33.7  & --00 11 49  &  $5.4\pm2.1$ & $10.2\pm3.87$ & 3182 & 0.03 & 0.9 & 19.06 & point-like & 0.516 &BL\\
40$^{}$ & 13 42 12.0  & +00 29 49  &  $5.2\pm2.8$ & $3.99\pm1.33$ & 7603 & 0.60 & 2 & 20.41 & extended & 0.570 &BL\\
41$^{d}$ & 13 42 11.7  & --00 23 32  &  $4.2\pm1.3$ & $7.76\pm2.5$ & 5379 & -- & -- & -- & -- & -- &--\\
42$^{}$ & 13 41 47.8  & +00 31 08  &  $2.2\pm1.6$ & $2.84\pm0.9$ & 12635 & 0.28 & 1.4 & 20.56 & point-like & 1.335 &BL\\
43$^{}$ & 13 41 42.9  & +00 12 39  &  $6.7\pm2.6$ & $4.5\pm1.25$ & 9093 & 0.07 & 1.1 & 19.83 & point-like & 0.788 &BL\\
44$^{}$ & 13 41 40.5  & +00 15 44  &  $4.3\pm1.9$ & $3.55\pm0.97$ & 12732 & 0.80 & 3.3 & 18.59 & extended & 0.254 &AL\\
45$^{d}$ & 13 41 38.1  & +00 27 01  &  $1.3\pm0.4$ & $2.44\pm0.79$ & 18460 & -- & -- & -- & -- & -- &--\\
46$^{}$ & 13 41 34.4  & +00 28 08  &  $2.39\pm1.19$ & $2.01\pm0.61$ & 20020 & 0.22 & 1.2 & 20.26 & extended & 0.676 &photo-z$^{e}$\\
47$^{}$ & 13 41 28.4  & --00 31 20  &  $14.3\pm3.4$ & $13.5\pm1.73$ & 11128 & 0.03 & 0.3 & 20.63 & extended & 0.619 &photo-z$^{e}$\\
48$^{}$ & 13 41 27.8  & +00 32 13  &  $3\pm1.8$ & $2.98\pm0.85$ & 14784 & 0.57 & 2 & 21.71 & point-like & 1.764 &BL\\
49$^{}$ & 13 41 27.2  & +00 14 14  &  $10.1\pm2.5$ & $6.06\pm1.17$ & 12444 & 0.01 & 0.4 & 19.45 & point-like & 1.698 &BL\\
50$^{}$ & 13 41 27.1  & +00 23 28  &  $2.39\pm1.1$ & $1.7\pm0.5$ & 24650 & 0.03 & 0.3 & 21.38 & point-like & 0.120? &photo-z\\
51$^{}$ & 13 41 23.8  & +00 22 05  &  $1\pm0.9$ & $1.65\pm0.51$ & 24035 & -- & -- & -- & -- & -- &--\\
52$^{}$ & 13 41 22.5  & +00 28 26  &  $2.7\pm1.19$ & $1.48\pm0.54$ & 21578 & 0.60 & 2 & 21.9 & point-like & 2.250 &photo-z\\
53$^{}$ & 13 41 21.6  & --00 13 51  &  $7.5\pm3.4$ & $4.42\pm1.33$ & 8937 & 0.47 & 3.9 & 19.24 & point-like & 0.736 &BL\\
54$^{}$ & 13 41 18.1  & --00 23 21  &  $14.1\pm2.7$ & $12.2\pm1.3$ & 17665 & 0.48 & 2.6 & 19.63 & extended & 0.423 &BL\\
55$^{}$ & 13 41 17.1  & +00 21 51  &  $3.7\pm1.19$ & $2.68\pm0.59$ & 23633 & 0.22 & 1.5 & 21.17 & extended & 0.721 &BL\\
56$^{}$ & 13 41 15.0  & +00 19 46  &  $1.6\pm1.1$ & $1.45\pm0.54$ & 20778 & -- & -- & -- & -- & -- &-- \\
57$^{}$ & 13 41 02.9  & +00 15 48  &  $5.4\pm2.2$ & $4.05\pm0.99$ & 13280 & 0.07 & 1.1 & 19.62 & point-like & 1.038 &BL\\
58$^{}$ & 13 40 56.5  & +00 31 58  &  $3.5\pm1.9$ & $3.76\pm0.95$ & 13371 & $<0.01$ & 1.6 & 16.9 & point-like & 0.000 &STAR\\
59$^{}$ & 13 40 50.2  & +00 15 54  &  $6.1\pm2.5$ & $5.03\pm1.16$ & 11240 & -- & 6.4 & -- & -- & -- &--\\
60$^{}$ & 13 40 45.2  & --00 24 02  &  $15.6\pm4.1$ & $9.48\pm1.61$ & 10164 & 4.03 & 3.5 & 20.18 & extended & 0.215 &photo-z$^e$\\
61$^{}$ & 13 40 38.7  & +00 19 19  &  $8.6\pm2.8$ & $8.8\pm1.43$ & 10983 & 0.11 & 1 & 19.62 & extended & 0.244 &NL\\
\hline
\multicolumn{12}{l}{$^a$ 2-8\,keV count rate in $\rm cts\ s^{-1}$}\\
\multicolumn{12}{l}{$^b$ 2-8\,keV flux in units of $\rm 10^{-14}\,erg\,s^{-1}\,cm^{-2}$}\\
\multicolumn{12}{l}{$^c$ total exposure time in the 2-8\,keV  band from the merged PN+MOS image in seconds}\\
\multicolumn{12}{l}{$^d$ count rate and flux is from MOS}\\
\multicolumn{12}{l}{$^e$ photometric redshifts are from the SDSS catalogue}\\
\multicolumn{12}{l}{$^f$ X-ray/optical properties indicate Galactic star but spectroscopy is not available}\\
\hline
\end{tabular}
\end{table*}

\section{X-ray spectral analysis}
 
\subsection{Hardness ratios}  

\begin{figure*}
\rotatebox{270}{\epsfxsize=13.0cm \epsffile{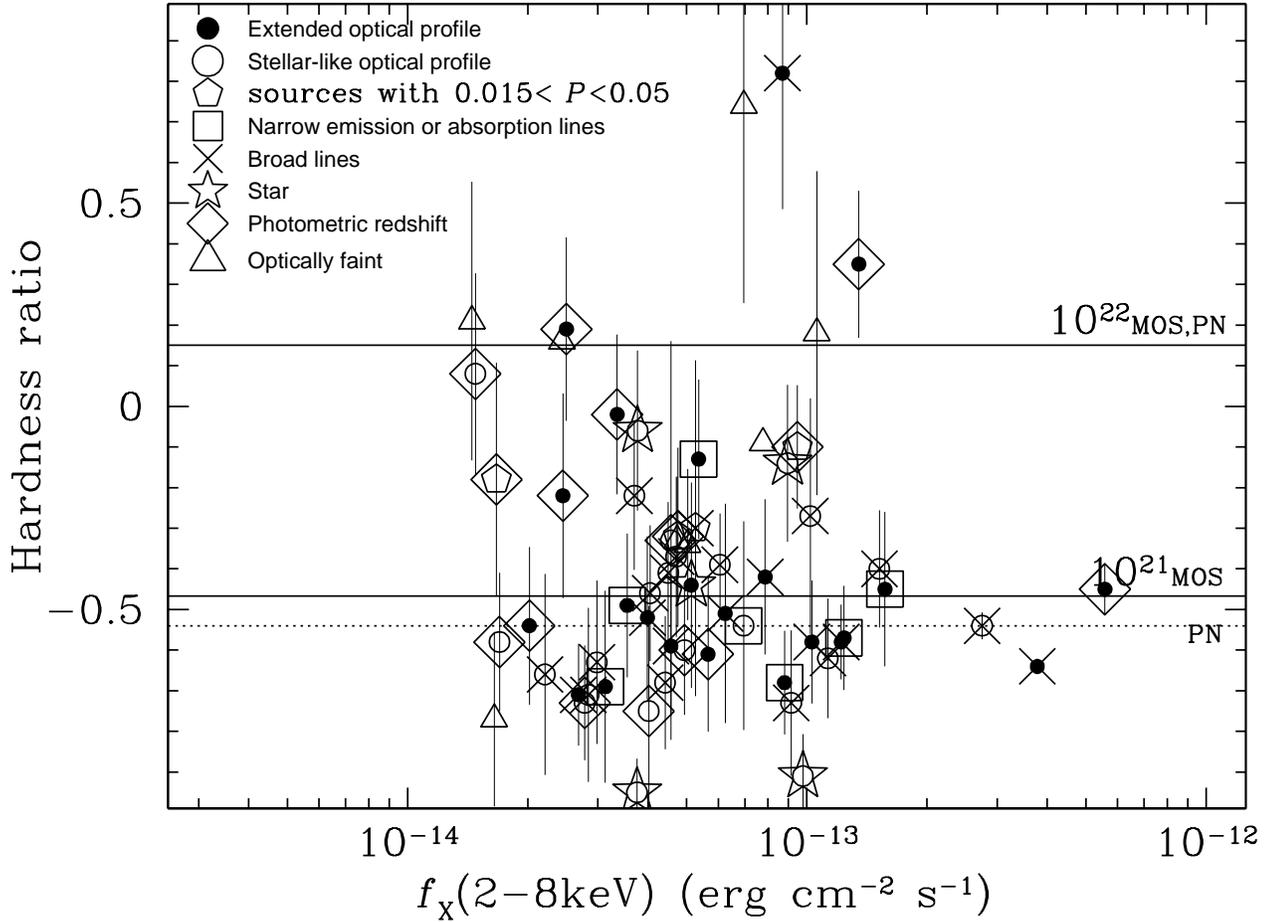}}   
\caption{The hardness ratio defined as $h-s/h+s$ (h and 
 s are the count rates in the 0.5--2 and 2--8\,keV bands 
 respectively)  as a function of the 2--8\,keV flux.
 For clarity we do not plot the errors for the sources with very large
 error bars and thus practically unconstrained hardness ratios (see
 text). Symbols as in Fig. \ref{colour}. Sources with no optical
 countepart that are absent from Fig. \ref{colour} are plotted as
 triangles. The horizontal lines denote spectra with a photon index 
 of $\Gamma=1.9$ absorbed by columns of $10^{21}$ and $10^{22}$ \cunits.   
 }
 \label{fxhr}
\end{figure*}

In Fig. \ref{fxhr} we plot the hardness ratio as a function of
flux. The column densities are derived from the  observed hardness  
 ratio using  the {\sc pimms}
 software assuming  a power-law  spectrum with $\Gamma=1.9$. 
Error bars correspond to the 68 per cent confidence level.  
For clarity we plot  the error bars only in the cases where the source
contains more than 5 net counts in both energy bands. 
As it was previously explained in section 3, 
 the hardness ratios are estimated from either the MOS or
 the PN detectors. As the MOS has a lower effective area at low
 energies, compared to the PN,  the same hardness ratio corresponds to
 different X-ray spectral properties (e.g. $\Gamma$, $\rm N_H$).
 
 In Fig. \ref{fxhr} we also plot
 the lines  corresponding to an absorbing column  of $\rm N_H=10^{21}$
 and $\rm 10^{22}\,cm^{-2}$, assuming a power-law spectrum 
 with a photon index of $\Gamma=1.9$ for the MOS and PN. 
 Note that the lines for
 the MOS and the PN are identical for high absorbing columns $\rm
 N_H>5\times 10^{21}\rm\, cm^{-2}$. 
 Although the sources span a large range of hardness ratios,
 most cluster around soft values: the average hardness ratio corresponds 
 to a spectrum $\rm N_H \sim 3\times10^{21}$ \cunits (for $\Gamma=1.9$).

\begin{figure}
\rotatebox{270}{\epsfxsize=6.5cm \epsffile{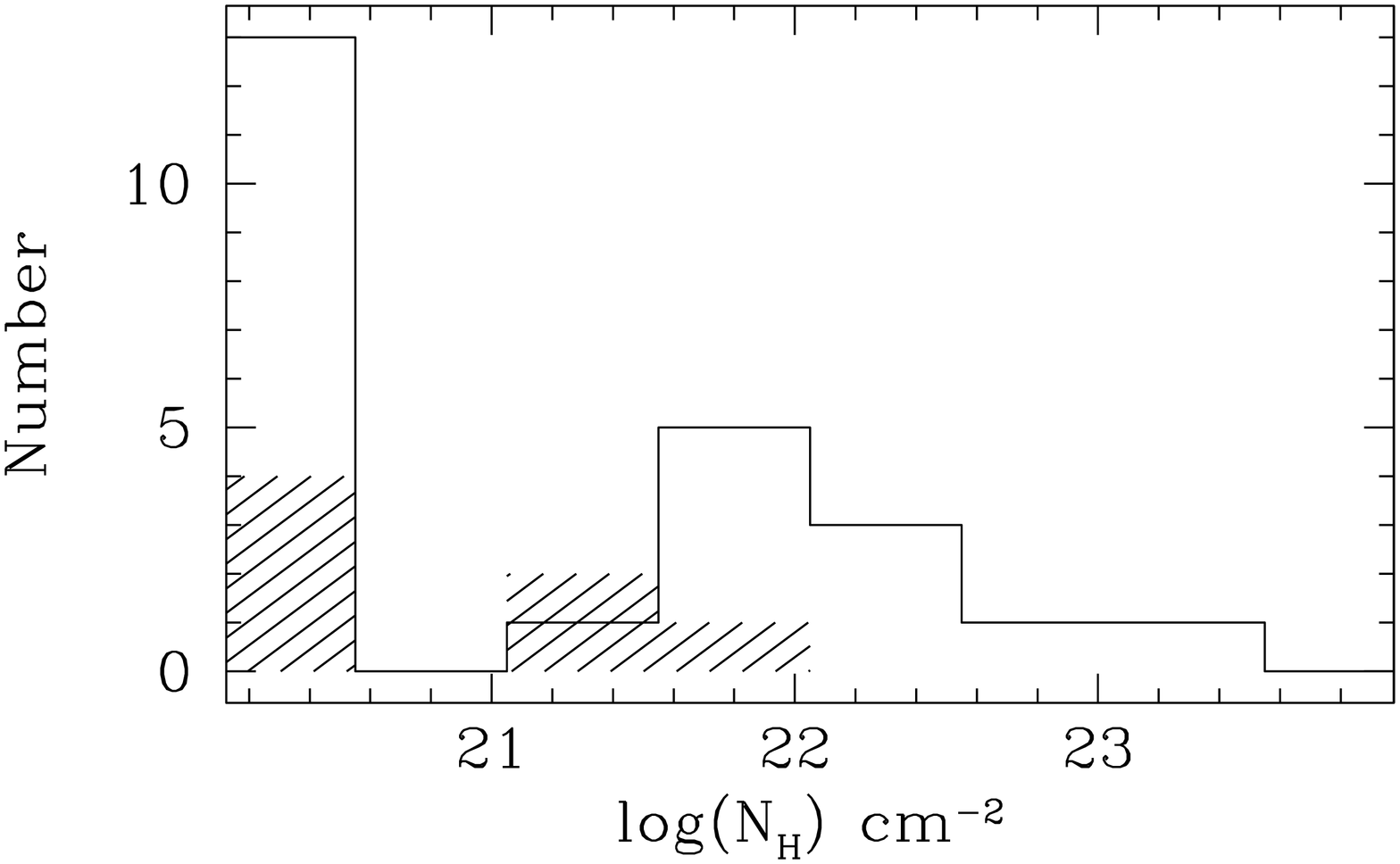}}
\rotatebox{270}{\epsfxsize=6.5cm \epsffile{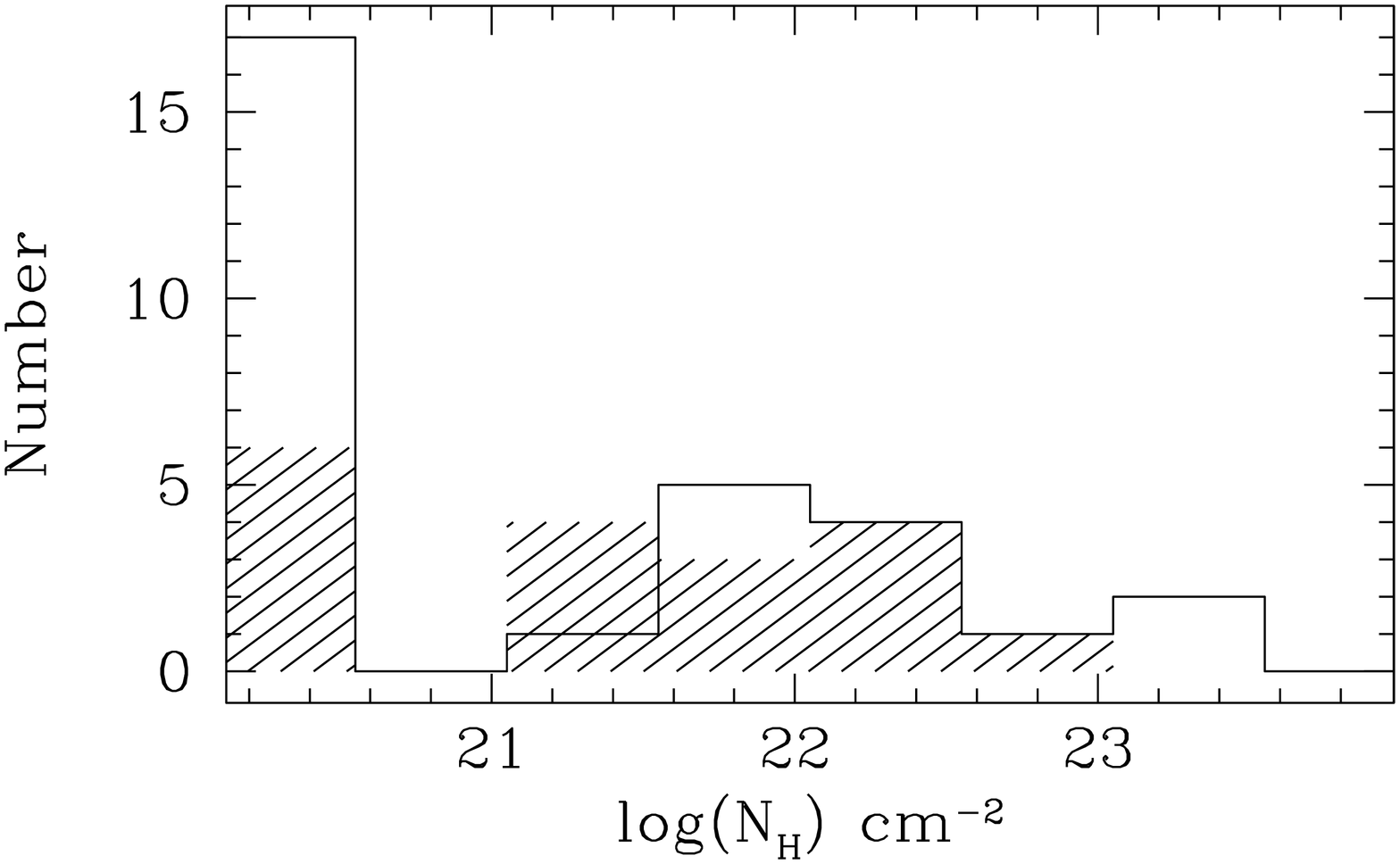}}
   
\caption{Upper panel: The  {\it rest-frame} column density distribution 
for the spectroscopically identified 24 BL (unshaded) and the 7 
 NL/AL (hatched) subsamples. Lower panel: The {\it rest-frame} 
 column density distribution for all AGN (including those with 
 photometric redshifts) i.e. the 30 BL and point-like sources (open) 
 and the  18 NL/AL AGN and extended sources (shaded).   
 }
 \label{nhcor}
\end{figure}

 The {\it observer's frame} column density  distribution 
 was presented in Fig.\ref{nh}.  
The column densities estimated there represent only a lower limit to
the true, {\it rest-frame} column densities.  The rest-frame column
density is  higher than the observed one as the k-effect shifts the
absorption turn-over at lower energies.  The relation between the
intrinsic rest-frame  and observer's frame  column density scales
approximately  as $(1+z)^{2.7}$ (Barger et al. 2002). The Galactic column
density of $2\times10^{20}\rm\, cm^{-2}$ has been subtracted before
applying the above correction.  In Fig. \ref{nhcor} (upper panel) 
 we show the {\it
rest-frame} column  density distribution for the 31 AGN which have
a spectroscopic redshift: the unshaded and  the
hatched histograms represent the BL and NL/AL AGN 
respectively. It appears that there is a small fraction of heavily
absorbed ($\rm N_H>10^{22}$ \cunits) BL AGN. 
 In the whole sample, there
are 11 sources whose 90 per cent lower-limit of the hardness ratio
corresponds to an observer's frame column density $\rm N_H>10^{21}$
\cunits. Two of them are spectroscopically identified as BL AGN
while one source is most probably a BL AGN  on the basis of its
stellar like optical morphology and high $fx/fo$ ratio.
Interestingly, the majority of the NL AGN present  soft hardness ratios
i.e. low column densities  consistent with the Galactic. However, as
the hardness ratio analysis provides only a rough  estimate of the
X-ray spectral properties, in the next section we attempt to constrain the column
densities via proper spectral fittings.

\subsection{Spectral fits to individual sources}
Here, we attempt to derive the spectral properties of a subsample 
 of interesting sources. More specifically, we derive the 
 spectral fits for the three QSOs which have high hardness ratio values
 (i.e. indicating hard spectra) as well as for all seven NL/AL AGN.    
For the X-ray spectral analysis the source counts are extracted using
a  radius of 18\,arcsec. The background spectrum is estimated from
image regions free from sources. Response matrices and auxiliary files
are generated using the {\sc sas} tasks {\sc rmfgen} and {\sc arfgen}
respectively. Spectral fits use  the {\sc xspec} v11.2 software. The
quoted errors correspond to the 90 per cent  confidence level.  

\begin{figure}
\rotatebox{270}{\epsfxsize=4.5cm \epsffile{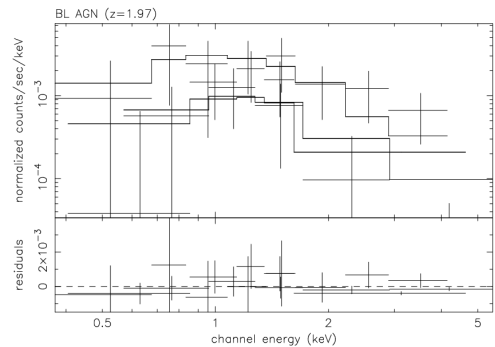}}   
\caption{The best-fit model and residuals to the 
 spectrum of a BL AGN which presents a significant 
 obscuring column (\#19 at z=1.97).}   
 \label{spectra}
\end{figure}

First, we attempt to constrain the intrinsic column densities of the 3 BL
AGN above, via proper spectral fitting,  using the {\sc xspec}
software. We perform the spectral fitting in the 0.3-8 keV band 
 where the instrument calibration is well known. 
 As our sources are faint, we use the
C-statistic  (Cash 1979) which does not require for the binning of the
data. However, the C-statistic does not allow for the 
 derivation of the goodness of fit probability  unlike the 
 $\chi^2$ statistic.  
We fit a power-law model absorbed by the Galactic column  ($\rm
N_H=2\times 10^{20}$ \cunits) and an additional intrinsic column
density.  The power-law slope is fixed to $\Gamma=1.9$ with the only
free parameters being the normalization and the intrinsic  column
density. The results are presented in Table \ref{individual}. Large
column densities are found in the   cases of the two spectroscopically
confirmed BL AGN: \#19 and \#10 having intrinsic columns of
$3\times 10^{22}$  and $\rm 10^{23}\, cm^{-2}$ at redshifts $z\approx
2$ and $z\approx 0.5$  respectively. 
 The first QSO presents UV excess having $\rm U-B \approx -0.6$ 
 while the second one has  $\rm U-B \approx +0.6$ and is extended
 in the SDSS images. In the third case, \#52,  a
candidate QSO with a photometric redshift z=2.25,  we find a 90 per
cent upper limit of $8\times 10^{21}$ \cunits for the intrinsic column
density.   
 In Fig. \ref{spectra} we present as an example the spectral fit
  to one of the  
 BL QSOs which present significant absorption (\#19 at z=1.970). 
 The spectrum is grouped so that it contains a minimum number of 5   
 photons per bin.  

In addition, we present the spectral fits to the 7 NL/AL AGN.  We fit
the same spectral model as above. The results are presented in Table
\ref{individual}. In most cases we find little evidence for large
rest-frame column densities ($>10^{21}$\,\cunits).
Object  \#30  at a redshift of $z=2.35$ 
 is associated with a NL QSO in the optical which has been 
detected in previous  \rosat and \asca surveys (Almaini et al. 1995,
Georgantopoulos et al. 1999). The \asca spectrum of this object can be
represented with a flat power-law  ($\Gamma \approx 1.5$) suggesting
some level of obscuration (Georgantopoulos et al. 1999). 
 Unfortunately, the \xmm spectrum has
 limited photon statistics,  failing to provide additional  clues on
 the nature of this object. Indeed, when we fit a power-law spectrum to 
 the data fixing the column density to the Galactic 
 we obtain $\Gamma= 1.77 \pm 0.36$ in agreement with 
 the \asca best-fit values.

\begin{table} 
\caption{Spectral fits to individual sources}
\begin{tabular}{llllll}
\hline 
Object & type & z & $\rm L_x^1$ & $\rm N_H^2$ & $\rm Counts^3$  \\
\hline 
 3 & NL & 0.370 & 44.0 &  $0.3^{+0.1}_{-0.1}$ & 153   \\
 9 & NL & 0.145 & 42.8 & $<0.03$  & 95   \\
10  & BL & 0.465 & 43.6 & $34^{+30}_{-20}$ & 39 \\ 
19 & BL & 1.970 & 45.0 & $3^{+3}_{-2}$ & 73 \\
21 & NL & 0.218 & 42.9 &  $<0.4$    & 56  \\
28  & AL & 0.240 & 43.6 & $<0.1$  & 140   \\ 
 30 & NL & 2.350  & 45.7 &  $<0.6$  & 58  \\
44   & AL & 0.254 & 43.1 & $<0.2$ &  107 \\
52  & BL & 2.250 & 44.7 & $<0.8$  & 34   \\
61 & NL & 0.244 & 43.6 & $<0.01$  & 223   \\
\hline
\multicolumn{6}{l}{$^1$ Logarithm of the absorbed Luminosity} \\ 
\multicolumn{6}{l}{in the 0.5-8 keV band (\lunits)} \\  
\multicolumn{6}{l}{$^2$ Intrinsic rest-frame column density in units
of $10^{22}$ \cunits} \\   
\multicolumn{6}{l}{$^3$ Sum of MOS and PN counts in the 0.5-8 keV band} \\   
\end{tabular}
\label{individual}
\end{table}

\subsection{The average source spectrum}
Next we present an analysis of the average source spectrum  in the
0.3-8\,keV band by co-adding the photons from all 61 sources in the
observer's frame. About 15 of the 61 sources either lie close to
gaps or bad pixels on the PN or outside the PN field of view. These
sources are still detected at a high confidence level on the  MOS
detectors which is less affected by CCD gaps. To improve the
statistical significance of our results in the analysis 
that follows we use the MOS data only. We note however, that using 
the PN detector does not modify our conclusions. 
The vignetting and PSF corrections were applied  by
creating individual auxiliary files for each source.  The auxiliary
files were then combined  using the {\sc addarf} task in {\sc
ftools}. The data have been  binned to give  at least 20 counts per
bin so that Gaussian statistics apply.We fit a single power-law model
to the data. The spectral fit results are presented in Table
\ref{coadded}.

The co-added spectrum of all 61 sources is best fitted by a power-law
spectral energy distribution $\Gamma=1.83^{+0.04}_{-0.05}$, much
steeper than the spectrum of the XRB at these energies.  
 This spectrum is dominated by several bright sources. Excluding the 
9 sources which contribute about 60 per cent of the counts  in the
0.5-8\,keV band we find $\Gamma=1.49\pm 0.06$. Note however, that the
exclusion of the  sources with  a large number of photons  may
introduce a bias against the sources with a soft spectrum.  This is
because of the large effective area  of the EPIC CCD detectors at soft
energies implying that a large  number of photons in the total
band  is most likely owing to a large  number of photons at
soft energies (0.5-2\,keV). 

The co-added spectrum  of the 24 spectroscopically confirmed BL
AGN gives $\Gamma=2.02^{+0.04}_{-0.05}$.
  Although  there are a few BL AGN
with significant photoelectric absorption,  
 as demonstrated above, their contribution to the  co-added 
 BL  AGN spectrum is minimal. 
  When we add the above 24 sources to the 6 
 (spectroscopically unidentified) sources which appear point-like 
 on optical images and present blue colours, and hence 
 are candidate QSOs, we obtain 
 $\Gamma=1.98^{+0.04}_{-0.06}$. 
The 7 NL/AL AGN have a flatter
spectrum $\Gamma=1.64\pm 0.11$. When we include the 
 11 nearby AGN (extended in the optical with red colours),
 we obtain a spectrum of $\Gamma=1.50^{+0.08}_{-0.08}$,
 similar to the spectrum of the X-ray background at these energies.
  Finally, the 8 'optically faint' sources i.e. those which 
 have no optical counterparts in the SDSS, have a very flat spectrum 
 (albeit with large uncertainties), $\Gamma\approx 1.1 \pm0.3$
 suggesting large amounts of obscuration.

\begin{table} 
\caption{Spectral fits to the co-added source spectra}
\begin{tabular}{cccc}
\hline 
Sample &  No & $\Gamma$ &\x2/dof   \\
\hline 
All    		               &  61   & $1.83^{+0.04}_{-0.05}$ & 235/238 \\
All excl. 9 bright sources     &  52  & $1.49^{+0.06}_{-0.06}$ & 126/132 \\
BL AGN    		       &  24  & $2.02^{+0.04}_{-0.05}$ & 247/151 \\
BL AGN (incl.photo-z)         & 30   & $1.98^{+0.04}_{-0.06}$ & 261/165 \\
NL AGN                        &  7   & $1.64^{+0.11}_{-0.11}$ & 37/32 \\
NL AGN (incl.photo-z)         & 18   & $1.50^{+0.08}_{-0.08}$ & 144/75 \\
Optically faint         & 8    & $1.09^{+0.30}_{-0.30}$ & 9/8 \\
\hline 
\end{tabular} 
\label{coadded} 
\end{table}

\section{DISCUSSION} 

The {\it XMM-Newton}/2dF survey provides the opportunity to 
explore in detail the sources which are responsible for a large
fraction of the XRB  (about 50 per cent
at the flux limit of our  survey $f_{\rm
2-8 keV} \approx 10^{-14}$\funits.)     
 
The hardness ratio distribution provides some initial information on
their X-ray spectral properties. 
  The population synthesis models (e.g. Comastri et al. 2001) 
 predict that about  two
thirds of the 2-10 keV X-ray selected sources with a flux higher than
$f_{\rm 2-10keV}>10^{-14}\rm \,erg\, sec^{-1}\, cm^{-2}$
should present a {\it rest-frame} column $>10^{22}$\cunits, 
 (roughly one third should have $\rm N_H>10^{23}$\cunits). 
 From Fig. \ref{nhcor} (upper panel), where the distribution of 
 the rest-frame column densities are plotted  for the 
 spectroscopically identified sources, it
appears that  a very small fraction of the sources  have
columns  $\rm N_H >10^{22}$\cunits. 
A deficit of strongly absorbed sources has also been noticed in the
bright \xmm samples of Piconcelli et al. (2002) and 
Caccianiga et al. (2004).    
 In Fig.\ref{nhcor} (lower panel) we plot the rest-frame 
 column density distribution for all sources 
 (including those with photometric redshift). 
 We see that only about 20 and 3 per cent of our sources 
 have a column density higher than $10^{22}$ and 
 $10^{23}$ \cunits respectively in disagreement 
 with the standard population synthesis models.

We further explore the spectral properties of the hard X-ray
sample as a function of X-ray flux. Even at the relatively faint flux
levels probed here the average spectrum of  all sources
($\Gamma\approx 1.8$) is much steeper than the spectrum of the XRB in
the 2-10 keV band $\Gamma\approx 1.4-1.5$ (e.g. Gendreau et al. 1995;
Miyaji et al. 1998). This may be due to a few bright  sources that
dominate the total counts thus contaminating  the average
spectrum. 
Excluding the 9 brighter sources  which contribute  about  
60 per cent of the
total source counts, we obtain a flatter spectral index  with
$\Gamma\approx 1.6$. 
The hardening of the  spectral index above, with the exclusion  of the 9
bright sources, could represent  a trend where harder (more absorbed)
sources are detected  with decreasing flux. 
However, when we divide our sample into two 2-8 keV flux 
bins we find no evidence for  spectral hardening.
 We obtain $\Gamma= 1.75\pm 0.06$ (49 objects)
and  $\Gamma=1.81\pm 0.06$ (12 objects) for the sources  with flux
$10^{-14}<f_{\rm 2-8 keV}<10^{-13}$ and
$f_{\rm 2-8 keV}>10^{-13}$ \funits respectively. 
The above conclusion is further  supported by inspection  of
Fig. \ref{fxhr} where the hardness ratios are  plotted as a function
of the 2-8\,keV flux. Apart  from the 3 brightest sources which
present  steep spectra there is no obvious evidence for the hardening
of the hardness ratio with decreasing flux. This is naturally 
 explained as the  absorbing columns are not large enough to 
 significantly affect the 2-8\,keV band
 (but can affect the soft 0.5-2 keV and thus the total 
 0.5-8 keV band, see e.g. Tozzi et al. 2001, Alexander et al. 2003).   
 Thus, there is no marked evolution of the
 hardness ratio in the 2-8 keV band, at least in the limited flux range probed 
 by our survey. 
 However, we note that at least two deeper X-ray surveys 
 appear to find evidence for a weak evolution 
 of the hardness ratio with the hard band flux 
 (Harrison et al. 2003, Fiore et al. 2003 but cf. Tozzi et al. 2001) 
 suggesting that a number of  sources with higher column densities 
 may be  detected at fluxes fainter than the limit of our survey. 
  
 As we have identifications for a large fraction of 
 our sources, we can explore the spectral properties 
 of individual classes of sources. 
 Out of the 61 detected sources 31 have  been spectroscopically
 identified with AGN. Among the subsample of hard X-ray sources with 
 redshifts we find (i) 24 BL AGN (ii) 7
 AGN exhibiting narrow emission or absorption lines. 
 The majority of these sources are BL
 AGN. As many of our identifications come from the 2QZ and the SDSS
 surveys which are primarily tuned  toward the detection of 
 UV excess QSOs, the large number of BL AGN found so far  among the
 spectroscopically identified sources is not surprising.
  For a large number of our sources we do not have spectra.
 However, we can obtain some  information on their nature 
 from their optical colours plotted in Fig. \ref{colour}. The 25 
 unidentified sources  comprise a mixture of (i) 6 point-like
 sources in the optical most of which are likely to be
 QSOs (ii) 11 optically extended sources  with
 primarily red colours suggesting low-$z$ galaxies  associated with
 nearby either obscured or low luminosity AGN (iii)
 8 sources with no optical counterpart.     
 
We attempt to constrain the co-added spectral properties of the above
populations to explore whether they are consistent with those of the
XRB. For the 24 BL AGNs the average spectrum is steep having a power-law
photon index $\Gamma=2.02^{+0.04}_{-0.05}$. 
The addition of the 6 probable QSOs for which we have 
 determined photometric redshifts yields again 
 a comparable steep spectrum $\Gamma\approx 1.98$.  
Detailed spectral fits of
individual sources reveal only two BL AGN with non-negligible
rest-frame absorbing column densities ($>10^{22}$ \cunits). Note that
although there have been many claims for the detection of BL AGN
at moderate to high redshift with significant absorbing columns  only
a limited number of cases are based on spectral fits (e.g. Reeves \&
Turner 2000; Georgantopoulos et al.  2003; Brusa et al. 2003).

 The average spectrum of the  7 NL/AL AGN is somewhat flatter than 
 that of BL AGN ($\Gamma \sim 1.6$). The addition of the 11 
 galaxies with red colours, to the above sample 
 results to an average spectrum of $\Gamma \sim 1.5$ similar 
 to the spectrum of the XRB in this band. 
 Interestingly, the individual fits to the spectra of the 7
 spectroscopically identified NL/AL AGN  do not exhibit
 significant excess X-ray absorption  above the Galactic. 
 This can be explained if some
 of these are relatively weak AGN where  the optical nuclear light
 is diluted  by the strong galaxy component (e.g. Georgantopoulos et
 al. 2003, Severgnini et al. 2003, Maiolino et al. 2003).  

Finally, the 8 'optically faint' sources, namely those which
 have no optical counterpart at the limit of the SDSS, present a very flat 
 spectrum, albeit with large uncertainty ($\Gamma\approx 1.1\pm 0.3$).  
Obscured QSOs at high redshift may be associated with the last subclass
of sources. Indeed, at high redshift the flux observed in the optical
band originates in  rest-frame UV  wavelengths, which are most sensitive 
to reddening. However, even if all 8 sources which have no optical
counterpart, were associated with  obscured  QSOs, this number would
still be too small  to be consistent with the 3/1 ratio of obscured to  
unobscured  AGN observed in the local Universe.

\section{conclusions}

 We are reporting on the X-ray spectral properties of the hard X-ray
 selected sample detected in the  {\it XMM-Newton}/2dF survey which covers about 1.6
 deg$^2$. 61 sources have been detected in the 2-8 keV band 
 down to a flux limit of $f\approx 10^{-14}$ \funits.
 At this flux, our sources contribute about 50 per cent of the XRB
 in the 2-8 keV band. Spectral identifications exist for 34 sources
 (these produce about 30 per cent of the X-ray background) 
 while  photometric redshifts have been determined for another 17   
 sources which are most probably AGN. 8 sources have 
 no optical counterpart to the photometric limit 
 of the SDSS survey ($r\approx 22.5$). The main conclusions can
 be summarised as follows:   

 (i) The majority (24) of the spectroscopically identified sources are 
 BL AGN. However, this may be attributed to the fact that a 
 fraction  of our spectral
 classification comes from the cross-correlation of our catalogue with the  
 2QZ and the SDSS surveys which primarily detect  UV excess
 QSOs. 7 sources are identified  as NL (or AL) AGN on the basis of the
 lack of broad emission lines in the optical spectra. 
 Among the sources with no spectroscopic classification, 
 6 are point-like with blue colours and are most probably  
 QSOs.  Another 11 are associated with  optically extended sources
 with primarily red colours, suggesting that these are most likely 
 nearby AGN.    

 (ii) The average spectrum of all sources is represented by a
 power-law with a steep photon index of $\Gamma\approx 1.8$. When we
 exclude the  9 brighter sources, in terms of the number of  counts in
 the total band, we obtain $\Gamma \approx 1.5$.   The
 hardness ratio distribution shows a deficit of strongly
 absorbed sources  ($\rm N_H>10^{22}$ \cunits) in disagreement  with the
 standard population synthesis models. 

 (iii) The average spectrum of the 24 BL AGN is steep ($\Gamma\approx
 1.9$). However, there at least two BL AGN which present large ($\rm N_H
 > 10^{22}$ \cunits)  amounts of absorption, as shown from the detailed
 spectral fits.  Although these present great interest for 
 the physics of AGN, it appears that they do not form 
 a substantial fraction of the QSO population. 

 (iv) The spectrum of the 7 NL/AL AGN,   
 is flatter with $\Gamma \sim 1.6$. When we add 
 to the above, the 11
 spectroscopically unidentified sources which 
 are optically extended and present red colours,
 and thus are also probably associated with nearby NL AGN,
 we obtain a similar spectrum $\Gamma\approx 1.5$.   
 
 (v) The spectrum of the 8 sources which have no optical counterpart
is very flat , suggesting that these may be associated with
 obscured AGN at high redshift.

\section{acknowledgments}
This work is jointly funded by the European Union
 and the Greek Government  in the framework of the programme
 ``Promotion of Excellence in Technological Development and Research'',
 project ``X-ray Astrophysics with ESA's mission XMM''.

 We acknowledge use of the 100k data release of the 2dF Galaxy
 Redshift Survey. The 2dF QSO Redshift Survey (2QZ) was compiled by
 the 2QZ survey team from observations made with the 2-degree Field on
 the Anglo-Australian Telescope.

 Funding for the creation and distribution of the SDSS Archive has
 been provided by the Alfred P. Sloan Foundation, the Participating
 Institutions, the National Aeronautics and Space Administration, the
 National Science Foundation, the U.S. Department of Energy, the
 Japanese Monbukagakusho, and the Max Planck Society. The SDSS Web
 site is http://www.sdss.org/. The SDSS is managed by the
 Astrophysical Research Consortium (ARC) for the Participating
 Institutions. The Participating Institutions are The University of
 Chicago, Fermilab, the Institute for Advanced Study, the Japan
 Participation Group, The Johns Hopkins University, Los Alamos
 National Laboratory, the Max-Planck-Institute for Astronomy (MPIA),
 the Max-Planck-Institute for Astrophysics (MPA), New Mexico State
 University, University of Pittsburgh, Princeton University, the
 United States Naval Observatory, and the University of Washington.

\end{document}